\documentstyle[prl,aps,multicol,epsfig]{revtex}
\input epsf.tex

\newcommand{\be}{\begin{equation}}
\newcommand{\ee}{\end{equation}}
\newcommand{\ba}{\begin{eqnarray}}
\newcommand{\ea}{\end{eqnarray}}
\newcommand{\ban}{\begin{eqnarray*}}
\newcommand{\ean}{\end{eqnarray*}}

\newcommand{\moy}[1]{\langle #1 \rangle}

\newcommand{\demi}{\frac{1}{2}}

\begin{document}

\title{Feats, Features and Failures of the PR-box}

\author{Valerio Scarani\\Group of Applied Physics, University of Geneva\\ 20,
  rue de l'Ecole-de-M\'edecine, CH-1211 Gen\`eve 4, Switzerland}

\maketitle

\begin{abstract}
One of the most intriguing features of quantum physics is the
non-locality of correlations that can be obtained by measuring
entangled particles. Recently, it has been noticed that
non-locality can be studied without reference to the Hilbert space
formalism. I review here the properties of the basic mathematical
tool used for such studies, the so called Popescu-Rohrlich-box, in
short PR-box. Among its feats, are the simulation of the
correlations of the singlet and of other non-local probability
distributions. Among its features, the "anomaly of non-locality"
and a great power for information-theoretical tasks. Among its
failures, the impossibility of reproducing all multi-partite
distributions and the triviality of the allowed dynamics.
\end{abstract}


\section{Introduction}

This article is a review of recent investigations about
non-locality in physics. Here, non-locality is taken in the
following precise sense:

{\bf Non-locality:} \emph{There exist in nature channels
connecting two (or more) distant partners, that can distribute
correlations which can neither be caused by the exchange of a
signal (the channel does not allow signalling, and moreover, a
hypothetical signal should travel faster than light), nor be due
to pre-determined agreement (because the correlations violate the
so-called "Bell's inequalities").}

Any physicist knows that the "channel" that distributes such
non-local correlations are "entangled quantum particles". The
historical path to the discovery of quantum non-locality has been
reviewed many times --- and one of the common points I share with
GianCarlo Ghirardi is that each of us has written a book for a
broad audience whose core is precisely this discovery
\cite{livregc,livrevs}. To appreciate the scope of the approach to
non-locality reviewed here, it is useful to stress that non-local
correlations are {\em an observed phenomenon}. As such, they have
some independence of quantum physics. This implies: First, even if
one day quantum physics is overcome by a more general theory,
non-local correlations shall always be part of nature; just as
apples go on falling on earth, be it because of their "quality",
because of a "gravitational field", or because of the "curvature
of space-time". Second, although to date entangled quantum
particles are the only physical systems we can use to distribute
non-local correlations, it is a priori conceivable that one will
discover other physical processes leading to the same result.

In short, it is meaningful to {\em approach non-locality
independently of the formalism of Hilbert space}. This topic has
acquired growing importance in the last two years, and is still
very active at the moment of writing. I shall concentrate here on
what has been done with the basic mathematical tool developed to
study non-locality, the so-called {\em PR-box}. This simple
mathematical object has achieved several feats, uncovers
interesting features of non-locality, but cannot be the universal
building block of non-local correlations as one might have hoped;
this failure paves the way for generalizations and deeper
investigations.

\section{The PR-box}

\begin{center}
\begin{figure}
  \includegraphics{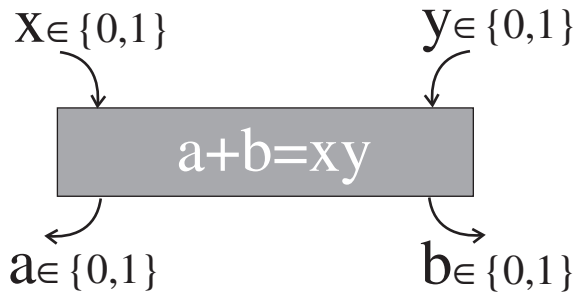}
  \caption{Pictorial representation of the PR-box.}
  \label{figpr}
\end{figure}
\end{center}

The PR-box has been named this way because it was introduced with
a physical meaning in 1994 by Popescu and Rohrlich \cite{pr94},
although actually its first appearance in the context of
non-locality dates back to mathematical works by Tsirelson
\cite{tsirelson}. It is a channel with two inputs ($x$ for Alice
and $y$ for Bob) and two outputs ($a$ for Alice and $b$ for Bob),
as schematically drawn in Fig.~\ref{figpr}. Each of the inputs and
the outputs is a bit, i.e. can take the values $\{0,1\}$; in all
what follows, all sums of two or more bits are to be taken modulo
2. The channel is defined by three requirements:\\ (I) The
no-signalling (ns) condition is satisfied:
\ba \sum_{b=0,1}P(a,b|x,y)\equiv P(a|x,y)&\stackrel{ns}{=}&P(a|x),\\
\sum_{a=0,1}P(a,b|x,y)\equiv P(b|x,y)&\stackrel{ns}{=}&P(b|y)\,,
\ea i.e. the marginals of Alice (Bob) don't depend on the input
used by Bob (Alice).\\ (II) Alice's and Bob's marginals are the
completely random distributions for both values of the input: \ba
P(a|x)\,=\,P(b|y)&=&\demi\,. \ea (III) Alice's and Bob's outcomes
are perfectly correlated according to \ba a+b&=&xy\,. \ea
Explicitly, this means that $a=b$ when either $x=0$ or $y=0$ or
both, while $a\neq b$ for $x=y=1$.

It is not hard to become convinced that no pre-determined strategy
can fulfill this rule \cite{gisin}, so the PR-box is a
no-signalling and non-local channel, just like quantum particles.
But in fact, this innocent-looking probability distribution is
more non-local than any correlation that can be distributed using
quantum states. To prove this, we can refer to the Bell-type
inequality derived by Clauser, Horne, Shimony and Holt
\cite{chsh}, that reads \ba
CHSH&=&E(0,0)+E(0,1)+E(1,0)-E(1,1)\,\leq \, 2 \ea where $E(x,y)$
is the correlation coefficient of the outcomes of Alice and Bob
for corresponding inputs. It is proven that quantum physics can
violate this inequality up to $CHSH_{QM}=2\sqrt{2}$ and not more.
Conversely, the PR-box reaches up to $CHSH_{PR}=4$, the largest
possible violation, since by definition $E(x,y)=1$ when one of the
inputs is 0, and $E(1,1)=-1$. In this perspective, it is natural
to ask, why can't this violation be reached using quantum physics.
The answer is not known, but some hints will be given in what
follows.

\section{Feats of the PR-box}

The PR-box slept as a mathematical curiosity for approximately ten
years. Its re-discovery was motivated by a result in the context
of "simulation of entanglement". This is the obvious starting
point for a review of the feats of the PR-box.

\subsection{Simulation of the singlet}

The first attempt of simulation of entanglement was the
hidden-variable program: can one reproduce quantum correlations
using only shared randomness? The negative answer given by John
Bell in 1964 is one of the main results of physics. It took almost
forty years to make the next step: if local resources are not
enough to simulate entanglement, what amount of {\em non-local
resources} should be supplemented? In quantum information, where
entanglement is a resource, this becomes natural question. The
first idea for a non-local resource was communication: we know
that communication is not a good physical explanation, but still,
it is something we can quantify, and thus provides a "measure of
non-locality". After some partial results we have reviewed
elsewhere \cite{methot} came the remarkable result of Toner and
Bacon \cite{toner}: to simulate the correlations of the singlet
(maximally entangled state of two qubits), it is enough to
supplement hidden variables with {\em a single bit of
communication}.

\begin{center}
\begin{figure}
  \includegraphics{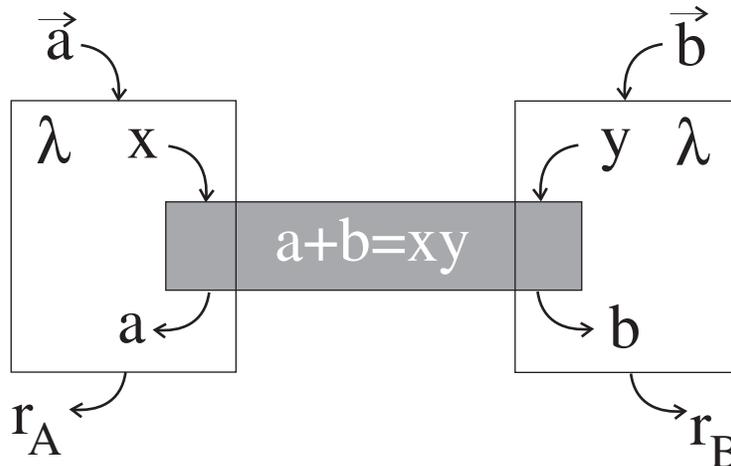}
  \caption{Schematic representation of the procedure for the simulation
  of the correlations of the singlet with a single use of the PR-box. See text for the explicit protocol.}
  \label{figsim}
\end{figure}
\end{center}

Inspired by this result, Cerf, Gisin, Massar and Popescu
\cite{cgmp} demonstrated that in fact the correlations of the
singlet can be simulated by supplementing hidden variables with
{\em a single use of the PR-box} (Fig.~\ref{figsim}). Although
mathematically based on the Toner-Bacon result, this work is a
major conceptual improvement: the PR-box is a strictly weaker
resource than a bit of communication, in particular because it
does not allow signalling. Since this is the triggering result for
the present-day interest in the PR-box, I give the procedure
explicitly, following Ref.~\cite{degorre}.

Alice receives $\vec{a}$, a unit vector representing the direction
along which her spin $\demi$ must be measured; she must output
$r_A\in\{+1,-1\}$. Similarly, Bob receives $\vec{b}$ and must
output $r_B\in\{+1,-1\}$. The goal is to achieve the statistics of
the measurement on the singlet state: $\moy{r_A}=\moy{r_B}=0$,
$\moy{r_A\,r_B}=-\vec{a}\cdot\vec{b}$. We suppose Alice and Bob
share a PR-box and local variables, in the form of unit vectors
randomly distributed on the sphere, which will be used in pairs
$(\vec{\lambda}_0, \vec{\lambda}_1)$. Here are the procedures
\cite{note0}:

\emph{Alice's procedure.} Alice verifies if
$|\vec{a}\cdot\vec{\lambda}_0|\geq |\vec{a}\cdot\vec{\lambda}_1|$;
if it holds, she inputs $x=0$ in the PR-box; it the opposite
holds, she inputs $x=1$ in the PR-box. From her output $a$, she
computes $\vec{\lambda}_A=(-1)^a\,\vec{\lambda}_x$ for the chosen
$x$. Finally, she outputs $r_A=
\mbox{sign}(\vec{a}\cdot\vec{\lambda}_A)$.

\emph{Bob's procedure.} Bob verifies if
$\mbox{sign}(\vec{b}\cdot\vec{\lambda}_0)=
\mbox{sign}(\vec{b}\cdot\vec{\lambda}_1)$; if it holds, he inputs
$y=0$ in the PR-box; it the opposite holds, he inputs $y=1$ in the
PR-box. From his output $b$, he computes
$\vec{\lambda}_B=(-1)^{b}\,\vec{\lambda}_0$. Finally, he outputs
$r_B= -\mbox{sign}(\vec{b}\cdot\vec{\lambda}_B)$.

To verify that this procedure gives the desired statistics, one
has to integrate over the uniformly distributed $\vec{\lambda}$'s;
see \cite{toner,degorre} for this last technical step.

\subsection{Simulation of other probability distributions}

A great hope stems out of the previous result: does the PR-box
play the same role in a general non-local formalism, as the
singlet does in quantum physics? Shared singlets allow to
distribute all quantum states: can all non-local distributions be
constructed by sharing PR-boxes? Although the final answer is
negative (as we shall discover later), the cases with positive
answer add to the feats of the PR-box. For instance:
\begin{itemize}

\item Any bipartite two-outcome correlation (i.e., $a$ and $b$ are
still bits but $x$ and $y$ can belong to larger alphabets) can be
simulated using PR-boxes (the number of instances is related to
some measure of complexity of the distribution) \cite{bp05,jm05};

\item Some multi-partite correlations which arise in quantum
physics can be simulated using PR boxes. More generally, with the
use of PR-boxes one can build correlations that reach the
algebraic maximum for a large family of multipartite Bell's
inequalities, thus generalizing the relation between the PR-box
and the CHSH inequality \cite{barrett,broad,india}.

\end{itemize}

For instance, one can simulate the correlations of the 3-party
Greenberger-Horne-Zeilinger (GHZ) argument \cite{ghz89} with local
variable plus one PR-box shared between (say) Alice and Bob
\cite{broad}. These correlations are (all the sums are modulo 2):
\ba
\begin{array}{lcl}
a_0+b_0+c_1&=&0\\
a_0+b_1+c_0&=&0\\
a_1+b_0+c_0&=&0\\
a_1+b_1+c_1&=&1\end{array} \ea where $a_x$ is the outcome of Alice
when her input was $x$, and so on. No shared randomness can
fulfill all four conditions. Suppose Alice, Bob and Charlie share
random variables $\lambda\equiv\{\alpha_x,\beta_y,\gamma_z\}$ that
fulfill the first three relations; then
$\alpha_1+\beta_1+\gamma_1=0$, contradicting the fourth
requirement. But it is easy to overcome this difficulty if we
provide Alice and Bob with a PR-box: Alice and Bob input $x$ and
$y$ in a PR-box; the outcomes of the box are $a$ and $b$. Alice
outputs $a_x=\alpha_x+a$, Bob outputs $b_y=\beta_y+b$, and Charlie
$c_z=\gamma_z$. By construction,
$a_x+b_y+c_z=\alpha_x+\beta_y+\gamma_z+xy$: the correlations of
the shared randomness are modified only when $x=y=1$, thus
fulfilling the fourth relation.

It is important to stress that this result and the analog ones
obtained for multi-partite distributions are much weaker than the
one about the simulation of the singlet: here, we require only to
be able to simulate the result of some very specific (albeit
highly interesting) measurements on quantum states.

\subsection{Monogamy of non-locality, and no-cloning}

\begin{center}
\begin{figure}
  \includegraphics[height=.12\textheight]{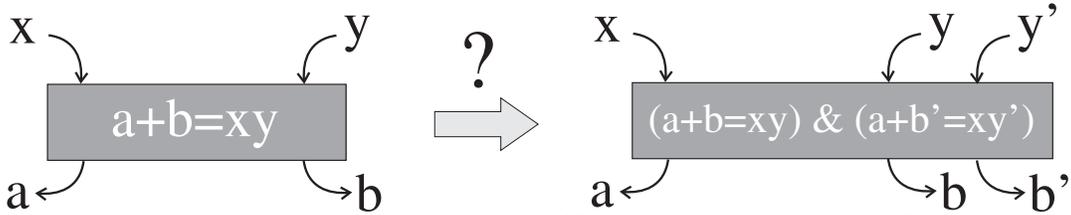}
  \caption{The transformation corresponding to perfect cloning of Bob's input-output channel.
  The new box allows signalling from Alice to Bob.}
  \label{figclone}
\end{figure}
\end{center}

The violation of the CHSH inequality is monogamous in quantum
physics: there exists no three-partite quantum state $\rho_{ABB'}$
such that both (say) $\rho_{AB}$ and $\rho_{AB'}$ violate CHSH
\cite{scarani}. The statement actually holds in general
\cite{mag}: there exists no no-signalling three-partite
probability distribution $P(a,b,b'|x,y,y')$, such that both
bipartite marginals $P(a,b|x,y)$ and $P(a,b'|x,y')$ violates CHSH.
This theorem (and its generalizations) imply in particular that
any no-signalling non-local theory contains a {\em no-cloning
theorem} \cite{mag,bar2}, generalizing the well-known tenet of
quantum physics \cite{clones}. In other words, the no-cloning
theorem can be demonstrated without any reference to the linearity
of Hilbert spaces.

The general proofs go beyond the scope of this paper, but we can
easily show the no-cloning theorem with the PR-box. Suppose that
Alice shares a PR-box with Bob ($a+b\,=\,xy$), and that Bob can
"clone" his part of the box in order to obtain another
input-output channel correlated with Alice according to
$a+b'\,=\,xy'$ (Fig.~\ref{figclone}). This cloning procedure opens
the possibility of signalling from Alice to Bob, because Bob can
compute $b+b'$ and this is equal to $x(y+y')$: since everything
but $x$ is known to Bob, he obtains the value of Alice's input.
Thus, exactly as is the case in quantum physics, perfect cloning
of boxes would lead to signalling.

Let's consider then an example of imperfect cloning, the one
described in Ref.~\cite{mag}. We look for the three-partite
no-signaling box $P(a,b,b'|x,y,y')$ such that $P(a,b|x,y)$ and
$P(a,b'|x,y')$ are the same distribution, and are "as close as
possible" to the PR-box. Let's assume the isotropic form \ba
P(a,b|x,y)&=&\frac{V}{2}\delta(a+b=xy)+\frac{1-V}{4}\,. \ea It is
easily verify that this distribution gives $CHSH=4V$. By the
monogamy of the violation of CHSH, we know that $P$ cannot violate
CHSH, whence $V\leq \demi$. On the other hand, $V=\demi$ can be
reached by a no-signalling distribution, because it can be reached
by quantum physics: the corresponding distribution is the one that
is obtained by applying a symmetric quantum cloning machine to
half of a singlet. Interestingly, the quantum cloning machine
which should be used here is not the universal one, but one which
clones optimally the states on the equator of the Bloch sphere
(the so-called "phase-covariant" cloner).

\section{Features of the PR-box}

The results reviewed above are clearly feats of the PR-box. Other
features have a more "neutral" character: we are content with
them, we'd have also been content if they had been different.

\subsection{Anomaly of non-locality}

In the last years, an "anomaly of non-locality" has been found: it
seems that, no matter which measure you use to quantify
non-locality, non-maximally entangled states are more non-local
than maximally entangled ones. All these results have been
reviewed recently \cite{methot}.

This anomaly manifests itself in particular when the amount of
non-locality is measured by the number of PR-boxes required to
simulate correlations obtained from quantum states. It has in fact
been proved that a single instance of the PR-box is {\em not}
sufficient to simulate non-maximally entangled states of two
qubits \cite{brunner}. I note here that the simulation of these
states is one the open problems of the whole field; the best
result to date \cite{ot} uses a signalling resource strictly
stronger than one bit of communication, therefore in particular
the procedure is asymmetric in time (Alice must do something
before Bob); and no simulation with no-signalling resources has
been found.

As mentioned, the appearance of the anomaly of non-locality with
PR-boxes cannot be considered as a failure, because all the
measures of non-locality known to date share this feature.

\subsection{Information-theoretical power}

Physicists are not the only community interested in PR-boxes:
theoretical computer scientists would be very glad to have one of
these magic boxes as a primitive!

Van Dam was the first to notice the power of the PR-box in an
information-theoretical context \cite{vandam}: he noticed that,
with such a resource, {\em communication complexity} would become
trivial. The scenario is the following: Alice and Bob receive each
a string ($\vec{x}$, $\vec{y}$) of $n$ bits; we require Bob to
output a bit $f(\vec{x},\vec{y})$. According to the definition of
$f$, this task requires some amount of communication. A
particularly hard case is the choice
$f(\vec{x},\vec{y})=\sum_{k=1}^n x_ky_k$: it can be proved (but it
is a quite intuitive result) that Alice must send Bob $n$ bits,
i.e. all her string, in order for him to give the correct output.
Moreover, no improvement is obtained when allowing Alice and Bob
to share entangled quantum states. However, the task becomes very
easy if Alice and Bob share a PR-box: they input sequentially all
the bits of their strings in the PR-box, and obtain outcomes $a_k$
and $b_k$ such that \ba
f(\vec{x},\vec{y})&=&A+B\,\equiv\,\sum_ka_k+\sum_kb_k\,. \ea Thus,
whatever the length $n$ of the string, Alice should send Bob just
one bit, $A=\sum_ka_k$. Van Dam suggests that the collapse of
communication complexity is an "implausible consequence": in other
words, that his result provides an argument to explain why quantum
physics does not reach up to the mathematical maximum of
non-locality. In fact, it is tempting to assume that the
non-locality of quantum physics is defined precisely by the point,
where communication complexity becomes non-trivial; at the moment
of writing, this is still a conjecture \cite{brassard}.

Apart from being useful for communication complexity, a PR-box
would also allow building {\em Oblivious Transfer (OT)}, an
important primitive of information theory \cite{ww}. This
primitive is defined as follows: Alice inputs two bits, $x_0$ and
$x_1$. Later, Bob inputs a bit $c$; he must receive $x_c$ as
output, while $x_{(c+1)}$ is forgotten. Note that OT is
signalling, therefore a signalling resource must be added to the
PR-box: here, a bit of communication. Specifically, the
implementation is: Alice inputs $x=x_0+x_1$ in the PR-box; she
gets $a$ and sends to Bob $m=a+x_0$. Bob inputs $y=c$ in the
PR-box, he receives $b=a+(x_0+x_1)c$. Upon receiving the bit of
communication $m$, Bob can compute $m+b = x_0+c(x_0+x_1)=x_c$.
Note however that this construction does not define
"unconditional" OT: it relies on the fact that Alice and Bob want
to use the OT to cooperate, or, in other words, that both partners
trust the other. This can be the case for the simulation of a
quantum channel \cite{ot}. In applications that involve an
untrusted partner, like {\em bit-commitment}, one cannot guarantee
that the other partner has not replaced the PR-box with a more
complex box \cite{bitcomm}. This is similar to the situation of
bit-commitment in quantum physics: it is possible only provided
one can trust that the other partner has not used ancillae to
enlarge his/her Hilbert space, but unconditional bit commitment is
impossible.

\section{Failures of the PR-box}

We have anticipated that the story of the PR-box is not only a
success-story: this nice mathematical tool cannot do everything
we'd like it to do. The failures are reviewed here. I don't
consider them as a reason for pessimism: rather, they indicate the
way to go for a deeper understanding of non-locality and of
quantum physics.

\subsection{Simulation of multi-partite distributions}

We had said above that the correlations of the 3-party GHZ
argument can be simulated by sharing bipartite PR-boxes
\cite{barrett,broad}. Other non-local multi-partite distributions,
however, cannot be simulated by bipartite boxes, even allowing an
arbitrary large amount of them \cite{bp05}.

Consider the following three-partite correlations for bits: \ba
\begin{array}{lcl}
a_0+b_1&=&0\\
b_0+c_1&=&0\\
c_0+a_1&=&0\\
a_0+b_0+c_0&=&0\\
a_1+b_1+c_1&=&1\,.\end{array} \label{corr3}\ea It is rapidly
verified that a GHZ-like argument holds for this distribution, so
it is non-local. The claim is that this correlation cannot be
simulated using bipartite PR-boxes.

Here is a sketch of the proof. Assume first that each pair of
partners share a single PR-box, that they can use only once.
Suppose now that Alice receives the input $x=1$: it can be the
third or the fifth relation. If it is the third, Alice must not
use $PR_{AB}$, the PR-box she shares with Bob: if she uses it, her
outcome will contain a random element, which is not compensated
because Bob's outcome is not taken into account --- and a random
element without correlations cannot give perfect correlations at
the end. In summary, to avoid mistakes, when Alice receives $x=1$
she can at most use $PR_{AC}$. But we can repeat the same
reasoning for Bob and Charlie, and we shall find that, when
everyone receives the input 1, no common PR-box can be used. In
conclusion, in order to avoid mistakes, the PR-boxes shall never
be used. The general proof is slightly more involved: basically,
the idea is that in a scenario with many PR-boxes and many
possible calls, there must be a last step: then, one applies the
argument we have sketched to the last step of the protocol and
reaches the same contradiction.

The origin of the problem is clear: we are considering
correlations, some of which involve only two out of three
partners. Thus, each partner cannot make up his/her mind, whether
he/she should use a given PR-box or not. Note that the
correlations (\ref{corr3}) cannot be obtained by measuring quantum
states. The simplest correlations known to date, which arises from
quantum physics and cannot be simulated with PR-boxes, are
five-partite and constructed on a similar basis \cite{bp05}: they
correspond to the stabilizer of the five-qubit cluster state, plus
a derived correlation that defines a GHZ-like paradox.

In summary, the PR-box is {\em not} the universal building block,
allowing to distribute all possible non-local correlations. This
negative result can be seen as ruling out a nonlocal-realistic
theory whose "elements of reality" would be local variables and
PR-boxes. In itself, this is not a dramatic problem: there is no
reason a priori why every non-local correlation should be
derivable from the simplest bipartite non-local box. Still, this
is in contrast with the role of the singlet state of quantum
physics: if $n$ partners share a sufficient number of singlets
pairwise, any quantum state can be distributed among them. How?
Well, by teleportation: one partner prepares the state on his own,
then applies the teleportation protocol to distribute it to the
others. This remark suggests that there is no analog of
teleportation for PR-boxes --- which is indeed the heart of the
matter as long as failures are concerned. The next paragraph is
devoted to it.

\subsection{Swapping and other dynamics}

Instead of focusing on teleportation, it is more natural to
consider whether the analog of {\em entanglement swapping} can be
defined. Take a first box defined by $a+b=xy$, and a second box
defined by $c+d=zw$. One can tentatively define a "coupler" to
apply to the extremities B and C, whose output is the box $a+d=xw$
shared between A and D (plus possibly some other outcome on which
to condition, the analog of the outcome of the Bell-state
measurement). There is no apparent inconsistency in this
definition: the new box is obviously no-signalling, and we don't
bother to find a "realization" of the coupler, it can be taken as
a new primitive of the theory.

However, it is natural to require that the coupler act
consistently, not only on the product of two PR-boxes, but {\em on
the whole set} of no-signalling probability distributions of the
form $P(a,b|x,y)\times P(c,d|z,w)$ --- just like in quantum
physics, where coherent measurements are defined independently on
the state they act on. This requirement turns out to be impossible
to fulfill \cite{short}. The proof is rather involved, consisting
basically of checking all possible couplers and verifying that no
one acts consistently in the whole set of probability distribution
(apart from the trivial coupler that amounts at just forgetting B
and C, giving completely uncorrelated A and D).

Thus, no analog of the Bell-state measurement (and more generally,
of coherent measurements) seems to exist for PR-boxes. A similar
conclusion can be drawn from the fact that one needs $\sim 2^n$
PR-boxes in order to simulate some tasks Alice and Bob can perform
by sharing $n$ singlets \cite{broad}. In fact, the exponential
gains in quantum information come from the possibility of
performing coherent measurements.

Other failures related to the dynamics have been discovered:
\begin{itemize}
\item No analog of the single-qubit Hadamard gate, and in fact of
most one-qubit rotations, can be defined on PR-boxes \cite{bar2}.

\item The correlations of the PR-box, and actually all those that
are more non-local than allowed by quantum physics, cannot be
derived from a reversible no-signalling unitary evolution
\cite{massar}.

\end{itemize}

\subsection{Entanglement vs boxes}

The way out of these failures is not known. In fact, we don't know
if one can define a no-signalling non-local theory with
"interesting" dynamics which is different from quantum physics,
and possibly out of which quantum physics could "emerge".

A few observations may be useful for a better understanding of the
difference between entanglement and non-local boxes. The non-local
boxes are built to describe the "measurement process": you put an
input and get a result; conversely, a quantum state can be defined
without assuming that it is going to be measured. In spite of
being non-local and intrinsically non-deterministic, PR-boxes are
"classical" channels, on which it has no meaning to define a
superposition principle or a coherent measurement.

There is something else. Entanglement cannot increase if the
partners apply local operations and if they communicate
classically; that is, classical information is a resource that can
be used along with entanglement, without (so to say) changing the
rules of the game. This is clearly not the case for non-local
boxes: the few possible local operations do not increase
non-locality, but classical communication is by definition
forbidden: it increases non-locality... and its signalling! This
remark can be summarized by saying that, in its present
formulation, the theory of non-local boxes is {\em poorer in
resources} than entanglement processing, because it does not have
a "free resource" to be used alongside with the boxes.

\section{Conclusion}

"[L'universo] non potrà essere letto finch\'e non avremo imparato
il linguaggio e avremo familiarizzato con i caratteri con cui \`e
scritto. E' scritto in linguaggio matematico, e le lettere sono
triangoli, cerchi e altre figure geometriche, senza le quali \`e
umanamente impossibile comprendere una singola parola." This is
Galileo's celebrated quotation from \emph{Il Saggiatore},
stressing that Nature is written in mathematical language. In this
text, Galileo proposes also the Alphabet for this language:
"triangles, circles and other geometrical figures". Four centuries
later, there is still a lively discussion about this Alphabet, the
building blocks of the description of Nature. Since we know that
Nature is non-local, one or more non-local objects must belong to
the Alphabet.

In this paper, I have presented the simplest non-local object, the
PR-box. Is it a letter of the Alphabet, or is there any other
non-local primitive with better properties, or is the mathematical
path sketched here bound to fail? Why is quantum physics
non-local, but {\em less non-local} than would be allowed a priori
by no-signalling \cite{pr94}? These are cards in God's play at
which we hope to sneak a look soon \cite{livregc}.


I thank Nicolas Gisin and Andr\'e M\'ethot for insightful comments
on a first draft of this review. I acknowledge financial support
from the Swiss NCCR "Quantum Photonics" and from the European
Project QAP (IST-FET FP6-015848).

\bibliographystyle{aipproc}   

\begin{thebibliography}{9}
\bibitem{note0} Two remarks: (I) Contrary to what happens in
quantum physics, the procedure presented here (and in fact all the
simulation procedures found to date) treats Alice and Bob in an
asymmetric way. (II) In Ref.~\cite{degorre}, $r_B$ is defined with
the opposite sign. This gives rise to the correlations
$\moy{r_A\,r_B}=\vec{a}\cdot\vec{b}$, that cannot be obtained by
direct measurement in quantum physics (the corresponding "density
matrix" would have negative eigenvalues on some entangled states),
but that can be easily produced in the lab by assuming that Bob
systematically flips his result.
\end{thebibliography}

\end{document}